\documentclass[12pt]{article}
\usepackage[english]{babel}
\usepackage{a4wide}
\usepackage{latexsym}
\date{}
\begin{document}

\title{\bf {\large{The  Uniqueness of  the  World}}}

\author{Luis J. Boya\footnote{Dedicated to Emilio Santos in
his 70th birthday. I have enjoyed discussions with Emilio on
physics for the last 40 years.} \\
\normalsize{Departamento de F\'{\i}sica Te\'{o}rica, Universidad de Zaragoza } \\
\normalsize{E-50009 Zaragoza, Spain} \\
\normalsize{luisjo@unizar.es}}

\maketitle

\begin{abstract}

We follow some (wild) speculations on trying to understand  the
uniqueness  of our physical world,  from the field concept  to
F-Theory. \\

KEY  WORDS: Unification of forces, Uniqueness, High-Road approach,
M and F Theories.  \\

\end{abstract}

\vfill \eject

\section{ Special Relativity}

Is the Universe we experience derivable ``a priori''? That is, are
there enough reasonable assumptions from which one can understand
the major features we observe by looking at the world around us?
Some physicists (e.g. S. L. Glashow) call this the ``top-bottom''
approach, or the High Road to do physics. Einstein since around
1920 thought so, but this view has been fairly unpopular. In this
essay we are going to explore some old and recent avenues with
this idea in mind. \\

We notice we live in three spatial dimension, with another
dimension, time, flowing forward irrespective of us; we change
places at will, but, alas, time is not under our command. In a
first view we approach space as a continuum, and associate
somewhat arbitrarily three real number to specify a position. It
is perfectly possible that in the future we dispense with the
reals and substitute for some other class of numbers (integers?).
In fact, we can argue that the reals can never really be
observable, so in a strictly positivistic approach they should not
be used. Besides positions, we observe motions; time becomes
entangled with space, and the simplest thing is to attribute some
real number as time to mark events. (Static) geometry gives way to
kinematics. \\

Experience determines soon a first limitation: velocities are not
to be arbitrarily large; special relativity comes in. The lesson
is, I believe, nature abhors infinities, so no observable can
attain infinite values (the same argument goes also against the
real numbers, of course; it also hints towards atomicity, see
below). Velocity $c$ plays the role of infinite velocity in the
following sense: for an ``observer'' moving at $c$, it passes no
time: it is like going with infinity speed in prerelativistic
physics. By continuity, time slows down when we run. Special
relativity makes a greater strain in our intuition than the
general theory (Dirac is one of few  who understood and
explicitely said that: the conceptual jump from space to spacetime
is bigger than from flat space to curved), because absolute
(maximal) velocity, instead of absolute time, is more
counterintuitive and paradoxical that, say, the geometry of space
being determined by its matter content. Maximal velocity leads to
relative time, and one of the big tenets of classical intuition
(and classical philosophy), the steady flowing of an universal
time, falls into pieces: time goes along
characteristically with the moving observer. \\

So everything physical is propagated at finite speed. Including
forces, of course. Which forces? Well, at least macroscopically we
observe two forces, due to weight and charges. The main feature of
both, we have learned painfully in the last hundred years, is that
they are {\it gauge} forces. That is, there is a geometric construction,
the gauge arbitrariness being close to the patch freedom or
coordinate choice in the geometry of manifolds: consistency in
overlapping patches leads to the gauge group, and the concordance
of the equations with gauge invariance leads to conservation laws,
the most conspicuous feature of both gravitation and electricity.
In return, this implies the field laws, as it gives freedom to the
propagating fields and, in retrospect, favours, if not selects,
the macroscopic dimension of space (-time): it is better to be in
a $(1, 3)$ manifold as regards only these two forces. \\

\section{ Electromagnetism}

Let us be more specific. Force with finite
propagation speed singles out a field description, as Faraday and
Maxwell claimed already in the XIX century: no newtonian
action-at-distance, but fields as propagating agents. The electric
field acts in a point of space, and charged matter responds by
moving under the influence of the local field; by reaction, the
charged matter distribution controls the existence and propagation
of the fields. The primitive idea (Newton) that particles act
among themselves with no further reality, is superseded in the
field approach, which acquires degrees of freedom of its own:
Maxwell predicted electromagnetic waves, one of the most
sensational discoveries in the history of science. \\

Let us call ${\bf F}$ the (electromagnetic, em) field, and ${\bf j}$ the charged
matter distribution; we emphasize ``charged'', because we learn soon
that matter, indeed most of the matter, is neutral (= globally
uncharged). The relation between ${\bf F}$ and ${\bf j}$ must be of contact type
(today expressed as local (differential) equations in the
numerical continuum; how will it be in the future?). The liberties
of the field manifest themselves in the differential character of
the equation: so, through the integration constant (or function),
the field exhibits its own degrees of freedom. Thus one of the
simplest formulation could be, with ${\delta }$ some first order
differential operator

\begin{equation}\label{eq:1}
\delta  {\bf F} = {\bf j}
\end{equation}

There are interesting solutions even when ${\bf j}  = 0$.This
formula is only half of the truth: it must be appended by stating
how does charged matter move in given fields (Lorentz force, see
below). If ${\bf F}$ and ${\bf j }$ are vector/tensor magnitudes,
the operator ${\delta }$ could be nilpotent, ${\delta^{2}} =0$:
technically, if they are $p$-forms, the only natural differential
operators are of square zero. Hence

\begin{equation}\label{eq:2}
\delta  {\bf j} = 0
\end{equation}

\noindent which is an invariance law: only conserved electric
matter can be coupled to the em field. As ${\bf j }$ is a
$p$-form, integration gives charge stability: $Q(t_0) = Q(t_1)$.
This is the distinguished feature of gauge theories, an automatic
conservation law. \\

Three important consequences can be drawn from this: first,
assuming ${\bf j}$ has the relativistic character of a $4$-vector,
(\ref{eq:1}) should be read as $3=3$ conditions, and not $4=4$,
that is, only conserved currents are to be coupled: conservation
of electric charge follows. Secondly, gauge invariance: the
Lorentz force, i.e. the relation by which ${\bf F}$ determines the
motion of charges, written in hamiltonian or lagrangian form, as
it should in any sensible formalism, must be in terms of a first
integral of  ${\bf F}$, and then the 1-form ${\bf A}$, such ${\bf
F} = d{\bf A}$, either postulated, or coming from $d{\bf F}=0$,
should enjoy some indeterminacy:  ${\bf A} \rightarrow {\bf A} +
d{\Lambda}$, where ${\Lambda }$ is the gauge function. This is the
patch arbitrariness we talked about above; in fact, in some
spacetime topologies ${\bf F}$ is always global, as curvature of
the connection, but ${\bf A}$ is not: ${\bf F}$ should be closed
but not necessarily exact. Properly, this gauge invariance e.g. in
the lagrangian formulation (which we omit) is the true reason of
charge conservation through the important Noether second theorem:
invariance under functional transformations imply relations among
the equations of motion, the fact that here we have $3=3$, not
$4=4$ equations. Third consequence, there is conservation of flux:
a single point charge $q$ should generate a field decreasing with
the $ 1/r^2 $ law, as flux conservation follows from (\ref{eq:1}):
a sphere no matter how large surrounding the charge should
encompass the same flux. \\

Therefore, no wonder about the $ 1/r^2 $ law, the same as Newton
discovered in gravitation, and for the same reason. Does this tell
us something about the geometry/dimensions of space? Well, it
makes the case for a preference for three extended {\it space}
dimensions: in one space dimension only, the Coulomb law will
determine confinement, there are not degrees for the field; also
in two dimensions (logarithmic potential), whereas in {\it four} space
dimensions the $ 1/r^2 $ law {\it for the potential} is scale invariant,
making the orbits unstable (besides complications at the quantum
level, like ``falling into the cente'', etc.). Worse for spacetime
dim  $ > 4$; we conclude the $d=3$ character of open space as being the
most satisfactory setting for an exact geometric force law. \\

The influence of the field on the motion of matter hinges on how
we describe matter; for a single structureless particle with mass
$m$, and charge $q$, the simplest equation would be (with ${\cal
F}$ the force and $u$ the $4$-velocity)

\begin{equation}\label{eq:3}
dp_{\mu}/ds = {\cal F} = q{\bf F}_{\mu \nu}u_{\nu}
\end{equation} \\

\noindent which is the Lorentz force. \\

\section{ Gravitation}

What about the geometry of this 3d space or 4d spacetime? As
Minkowski saw already in 1909, finite propagation speed hints
strongly to a 4d metric spacetime with signature (1, 3): the light
cone constitutes the set of light rays. Now we go back to Riemann
(whose physical insights are sometimes obscured by his gigantic
mathematical prowesses); in his geometric studies he already
hinted at a selfconsistent determination of the geometry of space
({\it not} of spacetime) on physical grounds, and therefore to
connect matter with curvature. It took more than fifty years for
this ideas to fructify, the main stumbling block being the
necessity, unthinkable in the pre-Maxwell times Riemann was
living, of uniting space and time. There should be another
geometric force which will shape the form of the universe for us.
Today it is only too clear how Einstein\'{}s General Relativity
program accomplishes this. We have the equivalence principle: a
test particle moves upon gravitation universally, depending not on
itself but on the surrounding matter: geodesic motion. This is
even simpler that the Lorentz force motion, because there is more
universality. \\

The next question is, how much deformation of space (curvature) is
produced by (a kilogram of) matter? It is hard to beat Wheeler\'{}s
{\it dictum}: space tells matter how to move, matter tells space how to
curve (bend). There is a $4d$ manifold ${\cal V}$ with a metric $g$, and then
with a geometric object, ${\bf G}$, coming from the spacetime structure,
necessarily some derivative(s) of the metric $g$, which should be
determined by the matter distribution, ${\bf T}$. So Einstein wrote

\begin{equation}\label{eq:4}
{\bf G} = \kappa  {\bf T}
\end{equation}

\noindent (to be compared with eq.(\ref{eq:1})) where the constant $\kappa $  converts matter to geometry.
But this equation refers
to all matter, because there are no ``unponderable'' matter. The
first sensible choice for ${\bf T}$ is to be a symmetric 2-tensor with positivity
requirements and conserved, $\delta {\bf T}=0$ (as opposed to the 1-tensor
${\bf j}$, as
there is electrically neutral matter, but not weightless matter).
Question arises, exactly as Einstein thought: how to extract, from
an arbitrary continuum 4-manifold ${\cal V}$, a 2-tensor with gauge
invariance (that is, with an automatic conservation law). The
struggle of Einstein along November 1915 to fix the right thing is
hectic; at the lagrangian level, Hilbert beated him for a week.
But there is nearly only a solution (good mathematics comes in
help of good physicists): if (${\cal V}, g$) is the $4d$ manifold, and {\bf Riem}
the (1, 3) Riemann curvature tensor, the only conserved tensor is
${\bf G}$, ( $\delta {\bf G }=0$ ), where

\begin{equation}\label{eq:5}
{\bf G} := {\bf Ric} - (1/2)g \ \ \ \mbox{where} \ \ \  {\bf Ric} =
\mbox{Tr}_1 {\bf (Riem)}
\end{equation} \

\noindent (We write the trace Tr$_1$ because by Tr$_2$ we mean
 Tr ${\bf Ric} =$ scalar curvature ${\bf R}_{sc}$: the object Hilbert wrote inside
 a lagrangian).And we repeat the scheme of above: first, there is conservation law,
$\delta {\bf T}=0$: only conserved matter can be coupled to spacetime curvature
(this point, automatic conservation laws has been forcefully
argued often by Wheeler). Second, equation (\ref{eq:4}) should read as $6
= 6$ numbers, not $10 = 10$ (the tensors are symmetric); the rest is
the gauge invariance (choice of patches to describe the $4d$ world).
Indeed, the lagrangian, as Emmy Noether taught both Hilbert and
Weyl, has a four-function invariance (under change of
coordinates), hence there are four constraints among the equations
of motion. \\

Third, the same $1/r^2$ law arises as before, at the level of the
geodesic motion $\gamma $, where $\delta \int_{\gamma } ds =0$,
closing into a beautiful circle, because the $1/r^2$ law was the
primordial discovered feature, already hidden in Kepler\'{}s first
law. Notice the paradox, seldom stated: for a single big mass $M$,
the metric far enough away falls as $r^{-1}$, so a closed orbit is
elliptical as in the $r^{-2}$ force law, but the curvature, which
plays the role of field intensity, should (and does!) fall off
faster, with $r^{-3}$ ; whereas the electric field of a point
charge falls off itself like $r^{-2}$: This is because geodesic
motion is inertial (no force); geodesics are fixed by the
connection (Christoffel symbols), falling like $r^{-2}$, not by
the curvature: electric forces are forces, but inertial ``forces''
are not. Is there then any role for the curvature? Of course: this
is the reason why the moon is more effective than the sun in tidal
motion on earth. \\

It is irresistible to add some comments to the magnitude of
Einstein creation (although, I repeat, less innovative that
special relativity). Definitive blow to the philosophical
standpoint that there is a empty space in which we add things to
fill it up:  space (and time) is not an ``a priori'' category: it is
created along with the matter in it. Rather, operatively speaking,
there is no space if there is no matter, and the latter fixes the
former; another blow to  a naive realism. In the search for an ``a
priori'' understanding of the world, many prejudices of naive
realism fall in the path of the scientific discoveries. \\

\section{Unification}

Are they really so different, electricity and gravitation? Both
are gauge theories, both universal or nearly so, both ruled by the
inverse square law… but in the face of it, very different also:
two signs for the charge, irrelevance of gravitation at the
microscopic realm. It had to be an outsider, an unknown young
physicist in the german city of K\"{o}nigsberg (at the moment occupied
by the russians) who dared to unite them. The most spectacular
feature of Kaluza\'{}s construction (1919) is to account for the
essential difference: negative/positive (and hence neutral)
electric matter {\it versus} just masses, and this by means of a
geometric device: let us add a circle at each $4d$ spacetime
point, and describe electric force as prompting motion along the
circle (two senses = two signs of the charge), whereas neutral
matter moves only in the other 3 large space directions. Even
today the boldness of Kaluza idea ranks, to us, close to
Riemann\'{}s first
inception of geometry as determined by physical effects. \\

To make a long story short, one would set the "gravitation"
equations in a circle times curved $4d$ space; the electric force
comes automatically, conserved, and with the two signs for the
charge. Besides, the disproportionate ratio of electric to
gravitant forces at the atomic scale sets the circle radius very
very small (O. Klein, 1926); compared even with the dimensions of
atomic nuclei, smaller for a factor of $10^{20}$. But there is a price:
a metric $g$ in ($D+1$) dimension gives rise, viewed from $D$-dim space,
to

\begin{equation}\label{eq:6}
g_{D+1} \rightarrow g_D + A_D + {\phi }_D
\end{equation}

\noindent that is, a $1$-tensor (em field potential) and an enigmatic $0$-tensor
(dilaton ${\phi}$), besides the metric $g$. \\

Both Einstein and Kaluza-Klein were far ahead of their times.
General relativity layed dormant for physicists until the
astrophysical discoveries in the mid-sixties, whereas the idea of
extra dimensions did not enter the mainstream of physics until the
\'{}80s, with many unbelievers still today. \\

Already in 1919 H.Weyl, thus before Kaluza, sought another electrogravitation
unification by allowing the scale of the metric $g$ to vary pointwise,
but this ``gauge'' invariance became much more useful as a point-dependent
phase in the charged fields. \\

\section{Other Forces}

Let us not stop at this. Nature is more versatile. Masses, charges
and light do not exhaust the physical world (inspite that Einstein
never cared about the rest, un {\it pecado de orgullo} we should
try to avoid). Could there be other forces, not ostensibly
expressible with the $1/r^2$ law, and still of geometric (gauge)
origin?  I confess I have no hint here but to retort to
experiments; it is only very difficult to understand atomicity, a
conspicuous feature of the natural world, unless there is some
quantum description (how, otherwise, does atomicity and
discreteness come from?), and some other forces.  Or, more
forcefully, arbitrarily large division of matter chunks is
unoperational: another {\it Irrlehre} to skip; from Democritus we
have learned this option. Indeed, granted the feebleness of
gravity forces at the atomic scale, evidence for nonelectric
forces in the microworld became incontrovertible in the late
20\'{}s and 30\'{}s: beta radioactivity and particles other than
electrons and protons in the nuclei,
hence nuclear forces of nonelectromagnetic origin. \\

The program then was clear: first, one should empirically state
the laws of these new forces, try to discover the intimate nature
(type of gauge, for example). There is a heroic struggle to
uncover the laws of both the strong and weak interactions, which
we forcefully omit, which lasted for about forty years, only to
state, at the end, the result: both are gauge forces, of course,
but both hide this character (none obeys effectively the $1/r^2$ law)
with mechanisms we do not fully understand even now: confinement
in one case ( the strong force), symmetry breaking in the other
(weak force). I cite here only two giants: Murray Gell-Mann (working
mainly in the weak forces found the gauge group for the strong) and
Steven Weinberg (working in the strong forces discovered
the broken gauge theory of the weak). Are then four forces too many? On the contrary:
to reduce all the observable phenomena to only four types of
interactions is a fantastic achievement. Yet not full unification,
as even the ``electroweak'' theory just juxtaposes two gauge theories
with an arbitrary mixing angle. The dynamics of the microworld seems to be
well described by the gauge forces associated to the group
$SU(3)_{QCD}\times (SU(2) \times U(1))_{electroweak}$. \\

But by now  Pandora\'{}s box is open, and we would like to apply
Kaluza\'{}s program to these new forces. We learned the lesson of the
fifth dimension in the same sense as Riemannn conception: the
space (-time) is physically determined, not only with respect to
shape, but we should strech the program towards understanding also the
dimension and the signature! Geometry should be as a whole
selfconsistently determined by matter. I do not see how people
totally happy with general relativity are so reluctant to accept
dynamically generated extra dimensions for our world; we feel only
four, yes, but we also feel matter as continuous and packed,
whereas it is discrete and space is practically empty ( indeed,
nuclear to atomic volume ratio close to $10^{-15}$). And following this
line, one should learn the lesson and go beyond: geometry is
primary to phenomena, so one must try to understand the extra
forces through geometry. In other words, there should be
constraints  in the extra dimensions, and fixed these, the extant
forces will follow. This should be the Top of the top-bottom
approach! \\

This program has not been fulfilled as yet; (it will be {\it el fin de
la  aventura}, the end of science. Inspite J. Horgan, we are still
light years away). So we pause for a moment and ask, instead, what
consequence will have to allow the dimensions to vary in order to
understand the forces, new and old, {\it more geometrico}. \\

\section{Extended Objects}

This is our understanding why extended objects are needed, or at
least very welcome, in a theory with more than five dimensions; it
is somehow antiintuitive, so we elaborate a bit. Since the times
of Democritus, a constant feature in physics has been to attribute
atomicity to matter; atoms should be the building blocks of the
Universe. But now, a more versatile point of view emerges: there
might be elementary {\it extended} objects which sould not be
operatively understood as made of little point particles. Now, why
the extended objects in the first place? I see a very neat
geometric/dynamical reason: in a universe with more than five
dimensions, and you need them if forces other that
electrogravitatory enter, the natural objects are strings or
membranes, or higher $p$-branes. Why? The key concept is duality
with gauge forces. Namely, electric forces, which we use for
abbreviation, come together with magnetic forces, as we know since
Amp\`{e}re and Oersted. Duality for particles (i.e. point particles)
occurs in $4$ dimension, but duality for strings occur in six, for
membranes in $8$, and for $p$-Branes in $2(p+2)$. The dual of a
string is a (magnetic) string only in six dimensions: a charged
particle in $4d$ describes a worldline, hence couples by the
Lorentz force to a $1$-form $A$ with a $2$-form $F$ as field
strength; $F$ dualizes to $\ast F$, another two-form. A string
spans a worldsheet, couples to a potential two-form $B$, with a
$3$-form field strength $G=dB$, dualizing to a Hodge partner only
in six dimensions. Similarly for higher
$p $-Branes. \\

Turning the argument around: if there are extra dimensions, to
find and deal with extended objects is just too natural. And,
{\it passim}, we overtake Democritus and Leucipus after 24 centuries:
elementary objects are not necessarily pointlike.\\

Here we hit a technical barrier, which makes one suspect a
theoretical scheme going beyond lagrangian quantum field theory is
called for: we do not know how to extract full exact consequences
of a string theory, not to speak of membranes or higher $p$-branes.
For $0$- and $1$-Branes the Nambu ( =extremal volume) and Polyakov ( =
general-relativistic) formulations are equivalent (although even
the nonperturbative regime in string theory escapes us). But for $ p > 1$
the minimal and relativity forms are inequivalent...
This is why, in our view, $M$-theory (see later) is stagnant. And we
did not adress yet another problem: granted there are extra
dimensions, and that we do not see them, why and how do they bend
(compactify)? Or rather, if all dimensions are finite (compact),
why ones are more compact than others?  \\

\section{SuperSymmetry}

But we still face other bigger problem, in the search for a unique
world. Already risen by Einstein, it is the ``marble versus wood''
problem. \\

The universe is structured like a building, with bricks held
together by a glue. The force fields are the glue, the fermions
are the bricks. Is this dichotomy fundamental? Quantum Mechanics
comes to stress the difference, making it more poignant: the
exclusion principle, which determines the existence of structures
in the Universe, does not apply to the coherent states of the
particles carrying the gauge forces, namely the classical fields.
Is there a description in which glue and bricks come together,
united but different? We tried unification, since Kaluza, for
interactions (glue). Now we ask for forces and matter itself. I do
not dare to say that supersymmetry ({\it Susy}) is compelling, but
certaintly looks as a provisional phenomenological step forward:
if there is a symmetry between bricks and glue, we ``understand''
why the two are necessary. So Susy comes to the fore, but from a
lateral entrance, however: the Susy partners of the extant
particles are new particles: in general, quarks and leptons go
with the fundamental representations of the gauge groups, whereas
the glues (gluons, photons) go with the adjoint. This were only
natural, for example, if the ``bricks'' have some solitonic character. \\

Supersymmetry extends to everything; simple quantum mechanics,
quantum field theory, electromagnetism (qed), gravitation, etc., have
more or less natural enlargements, becoming supersymmetric. There
is the so-called minimal extension of the standard model (MSSM),
with some phenomenological support (coincidence of the running
coupling constants at a very high scale, aprox $10^{16}$ GeV). The most natural
supersymmetry occurs because of the octonions: it is the triality
isomorphism between the three $8$-dim irreps of the Spin($8$) group;
in fact, this supersymmetry lies at the base of both superstring theory,
$11$-dim supergravity, and $M$-Theory. The three irreps
${\bf 8}_{v}, {\bf 8}_{L} \ \ \mbox{and} \ \ {\bf 8}_{R}$ realize $Spin(8)/Z_{2}$. \\

With troubles: we have not seen, so far, a single Susy partner of
extant particles/ force carriers. With advantages, in the sense of
that supersymmetry limits the dimensions and even the nature of
the extended objects. For example, superstrings live happy in ten
dimensions, but gravitons are even happier in eleven, with a
membrane floating around. So Susy seems a step forward in the
uniqueness program we address here. \\

It is worth recalling how, since 1974, the candidate for big
unification has changed. Once the electroweak theory passed the
first tests (neutral currents do exist; $W^{\pm}$ and $Z$ found
later), Georgi and Glashow proposed the first Grand Unification
model, based in the group SU(5). Insufficient, with neutrinos
massive. Then superstrings came along, with the veiled promise to
unify even gravitation in a convergent (renormalizable) scheme (1976).
Too early. In 1980 gravitation reached the climax together with
Susy, producing a momentaneous distraction (M. Duff) from
superstrings: eleven dimension supergravity is seen at our mundane
$4d$ world as a rich theory with $8$ gravitinos (for example).
{\it Embarras de richesse}? On the contrary, the gauge group is not big
enough to encompass the group of the standard model, which is a
geometric group: $O(1) \times  U(1) \times  Sp(1) \times  Aut_{0}(Oct)$,
or $Z_2 \times SO(2) \times SU(2) \times SU(3)$. Here $O(1)$ is $PCT$
and the last group is $SU(3) \subset G_2$,
the little group of octonion automorphisms. \\

Then in 1984/5 the superstring revolution came, with the promise of another T.O.E.
The main new ingredient was uniqueness: there are only five consistent superstring
theories, all living in $10=(1, 9)$ dimensions, all incorporating gravity, all
supersymmetric, some including two unique gauge groups... \\

\section{M-Theory}

By now we have uncovered the three ingredients I want to stress in
the program for a better understanding of the world around us:
extra dimensions, extended object and supersymmetry. Supergravity
and the different string theories fuse in the new, beautiful
$M$-theory (Townsend, Witten and Polchinski) who arose in 1995.
Beautiful but distant: after ten years, we are sill in the
begining of understanding it, no knowing its inner fabric. It is
fairly unique and rigid ( no coupling constants to adjust), but on
the face of it, not yet thoroughly satisfactory. \\

Ten years have barely passed, yes, and where do we stand? The initial
hopes of this $M$-theory did not materialize, and even the most
optimistic persons (Ed Witten) seem to have lost steam. At this point we
want to pinpoint some shortcomings of the present $M$-theory and
signal one potentially interesting new avenue. \\

First of all the gauge group of the standard model is not
reproduced; in fact the natural gauge group is too narrow, and
some string groups like $E_8 \times  E_8$ come ``out of the blue''.
Secondly, gravitation appears inextricably mixed with the other
forces, and one would like to separate it from gauge forces,
granted the insignificance of gravitation at the atomic scale. In
the third place, although $M$-theory is set with a maximum $11 = (1,
10)$ dimensions, strings live in ten and ourselves in four
uncompactified. We shall finish by enunciating $F$-theory, an
invention of C. Vafa (1996) which deserves serious considerations. \\

\section{F-Theory}

 According P. Ramond and B. Kostant (1999), the $11$-dim
Supergravity multiplet is related to the octonionic projective
(Moufang) plane $OP^2$, which is the exceptional symmetric space
$OP^2 = F_{4}/Spin(9)$; for example, the triplet: graviton ($+44$) -
gravitino ($-128$) + $3$-form ($+84$) comes from the Euler number ($3$) of
$OP^2$, generated, as irreps of $Spin(9)$, by the identity
representation of the group $F_4$. As remarked above, the particle
pattern does not naturally correspond to what we see (or expect to
see). Is there a natural extension, conserving the aesthetic
appeal, and enlarging the situation?  \\

C. Vafa advanced already in 1996 a possible extension (called
$F$-theory) in order to incorporate the $IIB$ string theory in the
same footing as the other string theories. But with the projective
insight of Ramond, there is a better formulation of this $F$-theory:
it is a theory living in 12 = (2, 10) dimensions, allowing for a
queer extended object (a ($2, \ 2$) supermembrane) and with a
temptative particle content expressable in the complex extension
(Atiyah) of the Moufang plane. Namely

\begin{equation}\label{eq:7}
Y = OP_{C}^2 = E_6 /O(10)\times O(2)
\end{equation}

(More precisely $Y := (E_6/Z_3)/[(Spin(10)\times U(1))/Z_2]$. The
particle supermultiplet has now $27$ members (Euler number of
$OP_{C}^2$), it is still supersymmetric, and offering more than
enough room (in fact, too much!) to encompass the experimental
(and some future!) multiplets. The two big groups above are
distinguished: $E_6$ is the best candidate for Grand Unification
(all nongravitatory forces) and $Spin^{c}(10)$ is the compact part
of $SO(10, 2) = Sp(2, Oct)$, the maximal extension of the Cartan
identity $B_2 = C_2$. The associated Ramond-Kostant multiplet has
now $2 \times 32768$ states (!), and neither graviton nor
gravitino(s) appear. It is too early to see how this gigantic
supermultiplet shrinks to the $128-128$ states of the $m=0$
multiplet of the Minimal Supersymmetric Standard Model. The new
scheme keeps most of the nice features as before, but
mathematically is simpler; for example, the supersymmertric
algebra in twelve dimensions contains just the ``de Sitter'' group
generators, presumably signaling the kinematic part, and then just
a selfdual sixform, which points to a special matter content.
Finally, the $27 = 3^3$ number for the multiplets makes one to
think both in the number of generations as well as the number of
colors. Only the future will tell us if Nature has chosen this
wonderful
and highly unique structure to describe our world.  \\

\end{document}